\documentclass[twocolumn,showpacs,preprintnumbers,amsmath,amssymb]{revtex4}

\usepackage{graphicx}
\usepackage{dcolumn}
\usepackage{bm}
\usepackage{amsmath}
\usepackage{amssymb}
\usepackage{subfigure}
\usepackage{psfig,float}
\def\OMIT#1{}
\begin{document}

\title{Flexible construction of hierarchical scale-free networks
with general exponent}

\author{J. C. Nacher, N. Ueda, M. Kanehisa, and T. Akutsu.}
\affiliation{Bioinformatics Center, Institute for Chemical Research, Kyoto University, 
Uji, 611-0011, Japan}

\date{7 May 2004}


\begin{abstract}
{\small{
Extensive studies have been done to understand 
the principles behind architectures of real
networks. Recently, evidences for hierarchical organization
in many real networks have also been reported. Here, we present 
a new hierarchical model which reproduces the main experimental 
properties observed in real networks: scale-free of degree distribution $P(k)$ 
(frequency of the nodes that are connected to $k$ other nodes decays as a power-law
$P(k)\sim k^{-\gamma}$) and power-law scaling 
of the clustering coefficient $C(k)\sim k^{-1}$.
The major novelties of our model can be summarized as follows: {\it (a)} The model 
generates networks with scale-free distribution for the degree of nodes with general exponent
$\gamma > 2$, and arbitrarily close to any specified value, being able to reproduce most of the observed hierarchical 
scale-free topologies. In contrast, previous models can not obtain values of $\gamma > 2.58$. {\it (b)} Our model has structural
flexibility because {\it (i)} it can incorporate various types 
of basic building blocks (e.g., triangles, tetrahedrons and, in general, fully connected clusters of $n$ nodes) and {\it (ii)} it allows
a large variety of configurations (i.e., the model can use more than $n-1$ copies of basic blocks of $n$ nodes). The structural 
features of our proposed model might lead to a better understanding of 
architectures of biological and non-biological networks.}} 
\end{abstract}

\maketitle

Recently, the importance of hierarchical modularity in the context
of biological networks 
\cite{hartwell, jerar1, review_gen} and some non-biological
networks \cite{jerar2, newman, vazquez} has
been pointed out and a number of theoretical models has been proposed. On the biological 
side, a major challenge is to understand
the relationships among fundamental elements such as
genes, proteins and chemical substrates in cells.
It is believed that some groups of interlinked elements 
(i.e., functional modules) can carry out relevant tasks in a
functional level \cite{hartwell}.
These functional modules can be integrated into larger groups,
generating a hierarchical organization \cite{jerar1}.
Though experimental work is much more important,
construction of adequate theoretical models is also important for 
better understanding of general principles behind architectures 
of biological networks.

Theoretical models for explaining real complex networks have 
evolved during the last years, 
from the classical random graph model \cite{erdos} and the small-world model \cite{watts} to scale-free network models 
\cite{bar1,review,amaral}. 
The most important feature of scale-free networks is
that the degree distribution $P(k)$ (frequency of the nodes 
that are connected to $k$ other nodes) decays as a power-law
$P(k)\sim k^{-\gamma}$.
In the earliest models of scale-free networks \cite{bar1,review},
probabilistic rules were employed to construct networks incrementally.
After that,
deterministic scale-free models introduced in \cite{determ1,determ2} were a step towards
simulation of a modular topology. However, these models 
lack the power-law scaling of $C(k)$, because their nodes 
have clustering coefficient $C_i(k_i)=0$. Recently, 
the modularity and hierarchical topology \cite{jerar1, review_gen, jerar2}
were introduced to explain all the observed properties in complex networks.
These observed properties of real networks with $N$ nodes can be
summarized as: 
scale-free of degree distribution $P(k)\sim k^{-\gamma}$, 
power-law scaling 
of the clustering coefficient $C(k)\sim k^{-1}$ and 
an independence of the network size $N$ and high value for the average 
of the clustering coefficient $C(N)$.
The clustering coefficient for each node $i$ is defined as
$C_i(k_i)=2n_i/(k_i(k_i-1))$, 
where $n_i$ denotes the number of edges connecting $k_i$ neighbors 
of node $i$, and
$C(N)$ reads as $C(N)=[ \sum_i C_i(k_i) ]/N$. Finally, the function $C(k)$ is defined
as the average clustering coefficient over nodes with the same degree $k$: 
$C(k)=[\sum_{i:k_i=k} C_i(k_i)]/N(k)$, where $N(k)$ is 
the number of nodes of degree $k$.


\begin{figure}[h]
\centerline{\protect
\hbox{
\psfig{file=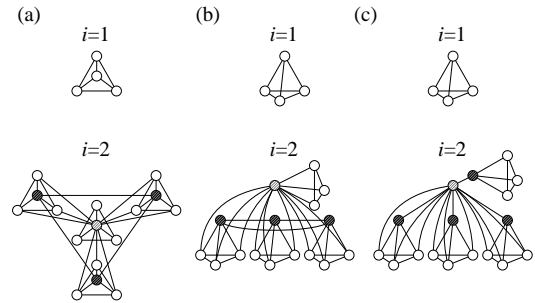,height=4.0cm,angle=0}}}
\caption{\small{(a) The RSMOB model \cite{jerar1}. Initial cluster with
four nodes, which are fully connected.
After the first replication the network
consists of 16 nodes ($4^2 = 16$). 
(b) The re-organized structure of (a)
to show clearly the similarities and differences between the RSMOB model and our proposed model. (c) Our proposed hierarchical model up to $i=2$.
We note that only one copy (among four copies) exists with one edge connecting to the 
main hub. The number of such copies is not 
restricted. When the number grows, 
$\gamma$ also increases.
}}
\end{figure}

In \cite{jerar1,jerar2} {\it Ravasz et al.} (the RSMOB model in what follows) suggested a hierarchical model 
to incorporate all the mentioned observed properties in the same framework.
The model starts with a fully connected
module of four nodes (the number 
of nodes in the initial module can be different), and 
four identical copies are created, 
obtaining a network of $N=16$ nodes in the first replication ($4^2=16$ nodes).
This process can be repeated indefinitely.
We illustrate the process in Fig. 1(a).
It is mentioned in \cite{jerar1} that
the model follows a power-law scaling for $C(k)\sim k^{-1}$ and holds
a scale-free topology $P(k)\sim k^{-\gamma}$, with $\gamma =
1 + (\ln 4)/(\ln 3) \simeq 2.26$. By modifying the number of nodes in the initial module, 
the value of $\gamma$ changes. However, the value is constrained 
to $2< \gamma  \leq 1+ (\ln 3)/(\ln 2) \simeq 2.58 $, which indicates a small range 
of possible applications.


In this article, we propose a new hierarchical model
which integrates the observed properties of real 
networks in a single framework.
The model can generate a scale-free topology with
exponent $\gamma > 2$, and arbitrarily close to any specified
value.
In addition, our model has structural flexibility
because it can incorporate various types of basic building blocks
(e.g., triangles, tetrahedrons),
which might lead to better understanding of architectures of
biological and non-biological networks.
\begin{figure}[h]
\centerline{\protect
\hbox{
\psfig{file=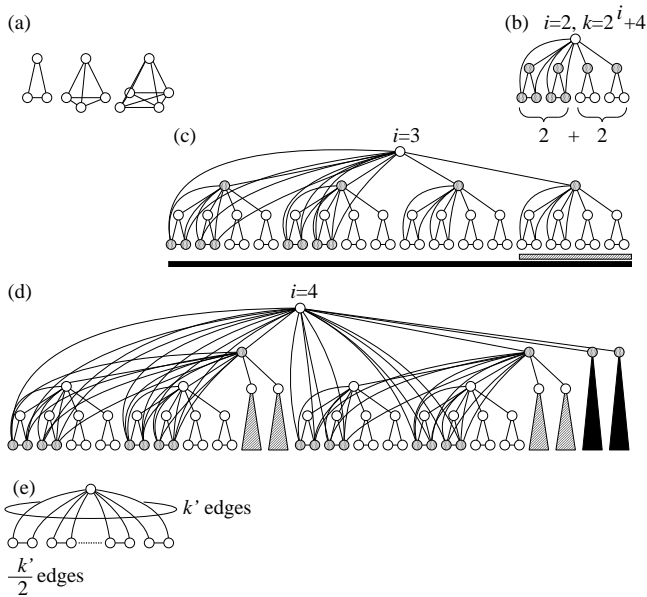,height=8.0cm,angle=0}}}
\caption{\small{Topology and construction of 
our proposed model. (a) The model can start with arbitrary number of 
nodes which are fully connected.
(b) Considering the initial cluster of three nodes, 
the two leftmost triangles have all 
their nodes connected to the main hub. 
This configuration is called the $(2+2)$ configuration.
The degree of the main hub is 
calculated as $k=2^i + 4$,
where $i$ is the number of iterations.
(c) Four copies of (b) are made, and
one node (the new main hub at this iteration) is added.
Fig. 2(c) contains four nodes as the second intermediate 
hubs. Each of these 
hubs holds $k$ edges, where
$k_j= [2^{j} + 4] +1$ and $j=2$.
(d) Following the same process, four copies of (c) are created. 
The process can be iterated
indefinitely constructing a network with power-law $P(k) \propto k^{-2}$. 
(e) Sketch of our model considering only the main hub with $k$ links 
and the nodes in the bottom level (i.e., non-hub nodes) that are connected
to the main hub. 
Since these non-hub nodes are connected by $k^\prime/2$ edges where $k^\prime=k-4$,
the clustering coefficient follows $C(k)\simeq 1/k$.
}}
\end{figure}

In order to explain an example of our model,
we look at the structure depicted in Fig. 2(b).
We see that there is a set of four triangles (fully connected 
cluster
of three nodes) with upper nodes connected to the main hub. 
In Fig. 2(a) we notice that the initial cluster could 
have different structures
and could be a fully linked initial cluster of 4, 5 nodes or
even larger number of nodes.
The initial cluster corresponds to the iteration of $i=1$.
Fig. 2(b) shows the iteration of $i=2$ where four copies
(the number of copies is selected according to the required $\gamma$) of 
the initial cluster are created and
one node in each initial cluster is linked to the main hub.
In addition, we note that only two out of the four
triangles have all their vertices connected to the main hub. 
For brevity,
we call a node in a copy corresponding to the main hub
in the $j$-th iterations an {\it $j$-th intermediate hub},
and call a node which is not the main hub or an intermediate hub
a {\it non-hub node}.
In Fig. 2(c), we show the network with iteration of $i=3$.
We make four replicas of the network in Fig. 
2(b) and connect the second intermediate hubs in
these copies to the main hub. 
The four non-hub nodes with the highest degree among the
non-hub nodes in two copies are also connected to the main hub.
In Fig. 2(d), we show the network with iteration of $i=4$
which is obtained by making four replicas of Fig. 2(c),
following the same process explained above.
This process can be iterated indefinitely.
The degree distribution of this network is 
dominated by the intermediate hubs. There is a main hub at the top of 
the structure and
new intermediate hubs appear
at each iteration. In Fig. 2(c) we see four 
nodes as the second intermediate hubs. 

Suppose that we have a network via $n$ iterations.
It is straightforward
to see that the degree of the main hub is $k=2^n+4$.
Since one edge is appended to the $j$-th intermediate hub
at the $(j+1)$-th iteration,
the degree $k_j$ of the $j$-th intermediate hub will be $k_j= (2^{j}+4)+1$,
if $2\leq j<n$.
We can see that the total number $N_j$ of $j$-th intermediate hubs
will satisfy $N_j=4^{(n-j)}$.
From $k_j=(2^j+4)+1$, 
we can write $\ln k_j \simeq j \ln 2$ and also from
$N_j = 4^{(n-j)}$,
we have $\ln N_j = (n-j) \ln 4 = c_1 -j \ln 4$.
From these expressions
it is straightforward to write:
$\ln N_j = c_1 + \ln k_j^{-(\frac{\ln 4}{\ln 2})} = c_1 + \ln k_j^{-2}$.
Hence, the number of hubs with degree $k$ (i.e., distribution of hubs with degree $k$)
in the proposed network follows the power-law $N_j \propto k_j^{-2}$. However, we must notice
that in a hierarchical network, the number of nodes with different degree $k$
is scarce, therefore the probability distribution of node degree is properly 
defined as $P(k)$=$(1/N_{tot})(N(k)/\Delta k)$, where $N(k)$ is the number of nodes 
with degree $k$, $N_{tot}$ is the total number of nodes,
and $\Delta k$ means that nodes are binned into intervals according to degree $k$.
In addition, we note that for the hierarchical model, $\Delta k$ changes linearly with $k$
(i.e., $\Delta k_{j+1}=k_{j+1} - k_{j}$=$2^j \simeq  k_j$). Hence, this linear dependence of
$\Delta k$ makes that the probability distribution follows in the proposed network 
the power-law $P(k)\propto k^{-3}$. In general, that binning gives rise to 
$\gamma =1 + \gamma^\prime$, where $\gamma^\prime$
means the exponent of the power-law distribution of hubs.

The construction can be generalized in the following way.
We denote by $(l+m)$-configuration one such that, 
at each (say the $i$-th) iteration,
$l+m$ copies of the network at the $(i-1)$-th iteration 
are created. With this configuration, we construct two types of connections between the copies and the main hub
at the $i$-th iteration:
connections between the $(i-1)$-th intermediate hubs and the main hub,
and connections between $l^{i}$ non-hub nodes with the highest degree
and the main hub. 

We notice here that this configuration is
flexible and can be modified.
There are two important and modifiable factors:
(i) the number of copies $(l+m)$ and the number of copies ($l$) 
for which some of non-hub nodes are connected to the main hub,
(ii) the basic building blocks (e.g, triangle, tetrahedron).
The former determines the value $\gamma$ and 
the latter affects the structure of network architecture.

Here we describe more about configuration of networks in our model. First we consider
a configuration which is able to reproduce the observed 
value of $\gamma=3.25$ in language network, which has a hierarchical organization \cite{jerar2}. This network is generated connecting two words to each other if they
appear as synonyms in the Merriam Webster dictionary \cite{jerar2}. We construct the network with the
($2 + 3$) configuration  ($k_j=(2^{j}
+5) +1$ and $N_j=5^{n-j}$ ), and we obtain 
$N_j\propto k_j^{-(\frac{\ln 5}{\ln 2})}$, where after binning we get 
$\gamma=3.3$. This value is in good agreement with the observed 
$\gamma=3.25$, which is not accessible with the RSMOB model. The reason is because the RSMOB model can only handle the case of $m=1$. 
Next we consider the general case. With ($l+m$) configuration,
we obtain $k_j=[l^{j} + (l + m)] +1$,
$N_j=(l+m)^{n-j}$, and $N_j\propto k^{-[\frac{\ln (l + m)}{\ln l}]}$, 
which indicates that by tuning the parameters $l$ and $m$ 
we have a network with exponent $\gamma$, which is arbitrarily close to
any required value above two.


\begin{figure}[h]
\centerline{\protect
\hbox{
\psfig{file=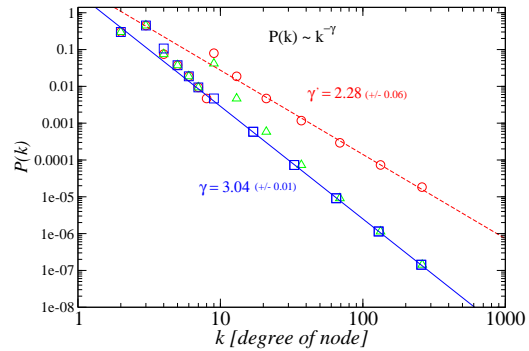,height=5.7cm,angle=-90}}}
\caption{\small{Circles: Distribution of nodes with degree $k$, $N(k)$, normalized to the 
total number of nodes, $N_{tot}$, (i.e., $N(k)/N_{tot}$). The network is constructed with the configuration $(2+2)$, up to $i=8$ 
and three nodes as initial cluster (triangles as building blocks). Dashed-line: Fit to the circles (only the main and intermediate
hubs). It shows a power-law with exponent $\gamma^\prime$ =2.28. Triangles: Probability distribution $P(k)$=$(1/N_{tot})(N(k)/\Delta k)$, 
where $\Delta k$ means that hubs with degree $k$ are binned 
into intervals $\Delta k_{j+1}=k_{j+1}-k_j=2^j\simeq k^j$ (i.e., $k_{j} < k\leq k_{j+1}$). We note that
for degrees $k$=8 and $k=9$ we used $\Delta k=2^1$. From $1< k \leq 7$, there are values for each $k$, and 
the binning is not required. Squares: Subtracting the value 5 in the axis of $k$ from the 
triangles (only the main and intermediate hubs). Continuous line: Fit 
to the squares. It shows a power-law with exponent $\gamma \simeq 3$.}} 
\end{figure}


From this construction of the hierarchical network we have several advantages if we compare with the RSMOB model \cite{jerar1}.
First, $\gamma$ can 
be arbitrarily close to any specified value above two, far from the restraints of the RSMOB model. Secondly, 
our procedure to generate the structure is more flexible and allows more variety of configurations. In Fig. 1(a) we show two iterations of 
the RSMOB model with 4 initial nodes, and in Fig. 1(c) we show our model up to $i=2$. Fig. 1(b) shows a re-organization of Fig. 1(a)
to point out similarities and main differences
between the RSMOB model and our proposed model. In the setup of Fig. 1, 
our model provides a dependence for the hubs
as $N_j \propto k_j^{-(\frac{\ln 4}{\ln 3})}$, and after binning we obtain
$\gamma=1 + (\ln 4)/(\ln 3)\simeq 2.26$, which is the same
result provided by the RSMOB model. In addition, we are more flexible with
our topology by increasing the number of copies. 
For example, with $(3+3)$ configuration, we obtain 
$N_j \propto k_j^{-(1 + \frac{\ln 2}{\ln 3})}$ and after binning we get
the value of $\gamma=2 + (\ln 2)/(\ln 3)\simeq 2.63$, which is not accessible 
with the RSMOB model \cite{comment2}.


Evidences for hierarchical organization in many real networks (biological and non-biological networks) have recently 
been reported. On the biological side, the metabolic network was analysed in \cite{jerar1,jeong, wagner} and 
the results showed that 
the value of exponent is $\gamma=2.2$, and the clustering coefficient $C(k)$ scales as $k^{-1}$. In \cite{wuchty2} 
protein domain networks were analyzed using data from different
domain databases and scale-free behaviors were reported
with values of exponents:
$\gamma=2.5$ (ProDom database), $\gamma=1.7$ (Pfam),
and $\gamma=1.7$ (Prosite).
A protein interaction network of {\it S. cerevisae}
was studied in \cite{wagner2} and it was found that $\gamma=2.5$.
In \cite{sitges}, the hierarchical signature of this network was
revealed showing that $C(k)$ scales as $k^{-1}$. From non-biological networks, we can also 
find some examples which hold a scale-free
topology integrated in the hierarchical organization \cite{jerar2}.
Here, we only mention
the type of network and the corresponding value of $\gamma$:
$\gamma=2.3$ for actor network \cite{bar2}, 
$\gamma_{out}=2.45$ and $\gamma_{in} =2.1$ (denoting the out and in-degree
distribution respectively) for {\it World Wide Web} \cite{bar2}, 
$\gamma=2.1 \sim 2.2$  \cite{greek} for Internet at the AS level
(interdomain level), 
and $\gamma=3.25$ for language network \cite{jerar2}.
In all these cases 
the scaling of $C(k)$ suggests the hierarchical organization \cite{jerar2}. For these examples 
with $\gamma>2$, our 
model is able to generate the scale-free topology with exponents 
arbitrarily close to the values shown above.




\begin{figure}[h]
\centerline{\protect
\hbox{
\psfig{file=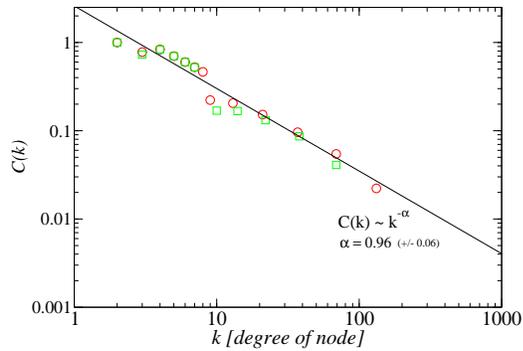,height=5.7cm,angle=-90}}}
\caption{\small{The clustering coefficient $C(k)$ evaluated with the configurations [$2+2$, up to 6 iterations] (circles) and [$2+3$, up to 5 iterations] (squares). In 
both cases, the building blocks are triangles. }} 
\end{figure}

In Fig. 3 we show the degree distribution of our model
with $(2+2)$ configuration, up to $i=8$.
As we explained before, the tail of that distribution (hubs) should follow a
power-law. Dashed-line indicates
one which fits the degree of the hubs of
our generated network. The meaning of this line is just distribution of nodes normalized 
to the total number of nodes. We see that the value of $\gamma^\prime$ is slightly different from
the theoretical value of 2, but the difference comes
from the approximation made from  $k_j=(2^j+4)+1$ to $\ln k_j \simeq j \ln 2$.
If we plot the dots after subtracting 5 units in the axis 
of $k$ and we fit
them, we could find exactly $\gamma^\prime=2$,
indicating that the difference between both results was coming from that
approximation. 
However, we are interested in the probability distribution of node degree 
$P(k)$=$(1/N_{tot})(N(k)/\Delta k)$. In Fig. 3 we show the probability distribution (triangles) 
after binning is applied for the hubs. In addition, we plot 
the probability distribution of the hubs after subtracting 5 units in the axis of $k$ (squares).
The continuous line is fitted to the squares and it shows a 
power-law probability distribution with exponent $\gamma = 3$. 

It is worth noticing that we can also reproduce the distribution without explicit construction
of the network. If we compute the values of $2^j+5$  (degree of hubs) versus
the values of $4^{(n-j)}$ (the number of copies) for $j=1,..,n$ and $n=20$, we can obtain
the power-law corresponding to $\gamma^\prime=2$ for the distribution of nodes and $\gamma=3$ for the probability distribution
after binning. It indicates 
that by generating a larger number of iterations in our model we 
are able to obtain exactly the predicted exponents. 


In Fig. 4, we calculate $C(k)$ for the $(2+2)$ and $(2+3)$ 
configurations in our model and we see the
power-law scaling of $C(k)\sim k^{-1}$,
which is also a key feature of the hierarchical network.
In Fig. 1(e) we show
a sketch of our model considering only the main hub with
$k^\prime$ ($k^\prime = k-l-m$) edges to non-hub nodes.
It is seen that there are $k^\prime/2$ edges among the non-hub nodes. 
From this, it is straightforward to see 
that the clustering coefficient for non-hub nodes
is: $C(k) = (k^\prime/2)/[(k(k-1))/2]
\simeq 1/k$,
showing the power-law scaling for the degree of clustering in our model. Concerning
the average of the clustering coefficient $C(N)$, its behavior in our model is independent of the network size $N$ as a consequence
of the power-law scaling of $C(k)$ \cite{ourwork}, in agreement with the observed properties in metabolic networks \cite{jerar1}.

It is interesting to note that our model holds a similarity with the model in \cite{bar1,bar2}, in particular with the 
preferential attachment feature. In that model, new nodes are being added in time step $t$, and the probability that the new node is
connected to an already present node $i$ depends on 
the degree $k_i$ of that node
$(k_i/\sum_j k_j)$. As we can see in Fig. 2, in each iteration 
we are adding a new node (main hub) plus copies of previous structures. The new hub is connected deterministically to 
the nodes in the non-hubs but only to those ones which have higher degree \cite{comment}. In that sense, a remanence 
of the preferential attachment 
concept is held in our model though the degree distribution for the non-hub nodes
does not follow the power-law as in the RSMOB model.

In conclusion, we have presented here a new model to reproduce the main features of the hierarchical organization, which is one 
of the central challenges in the field of network science. Our model holds
important properties as structural flexibility and its more general 
 capability to generate values of $\gamma > 2$, being able to reproduce most of the observed scale-free topologies, even in networks 
 with exponents above $\gamma=2.58$, where the RSMOB model \cite{jerar1} fails.
Therefore, our model might be a useful tool to uncover
the hierarchical features in biological and non-biological  
networks in a broader scope.

This work 
was partially supported by Grant-in-Aid for Scientific Research on Priority Areas (C) ``Genome Information Science''
from MEXT (Japan).





\begin{thebibliography}{99}



\bibitem{hartwell} L.H. Hartwell, J.J. Hopfield, S. Leibler and A.W. Murray, Nature {\bf 402}, C47 (1999).

\bibitem{jerar1} E. Ravasz, A.L. Somera, D.A. Mongru, Z.N. Oltvai, A.-L. Barab\'{a}si, Science {\bf 297}, 1551 (2002).

\bibitem{review_gen} A.-L. Barab\'{a}si and Z.N. Oltvai, Nature Genetics Reviews {\bf 5}, 101 (2004).


\bibitem{jerar2} E. Ravasz, A.-L. Barab\'{a}si, Phys. Rev. E {\bf 67}, 026112 (2003);


\bibitem{newman} M. Girvan, M. E. J. Newman, Proc. Natl. Acad. Sci. U.S.A. {\bf 99}, 7821 (2002).

\bibitem{vazquez} A. Vazquez, R. Pastor-Satorras, A. Vespignani, Phys. Rev. E {\bf 65}, 066130 (2002).

\bibitem{erdos}  P. Erd\"{o}s, P., A. R\'{e}nyi,  Publ. Math. Inst. Hung. Acad. Sci. {\bf 5}, 17 (1960).

\bibitem{watts}  D.J. Watts, S.H. Strogatz,  Nature {\bf 393}, 440 (1998).

\bibitem{bar1}  A.-L. Barab\'{a}si, R. Albert,  Science {\bf 286}, 509 (1999).

\bibitem{review} R. Albert, and A.-L. Barab\'{a}si,  Review of Modern Physics {\bf 74}, 47 (2002).

\bibitem{amaral} L.A.N. Amaral, A.Scala, M. Barthelemy, H.E. Stanley, Proc. Natl. Acad. Sci. U.S.A. {\bf 97}. 11149 (2000).


\bibitem{determ1} A.-L. Barab\'{a}si, E. Ravasz and T. Vicsek, Physica A {\bf 299}, 559 (2001).

\bibitem{determ2} S. Jung, S. Kim and B. Kahng, Phys. Rev. E, {\bf 65}, 056101 (2002).


\bibitem{bar2}  A.-L. Barab\'{a}si, R. Albert, H. Jeong, Physica A {\bf 272}, 173 (1999).




\bibitem{comment2} It is worth noticing that although our proposed model is much more 
flexible than the RSMOB model, our model can not reproduce
hierarchical networks with $\gamma < 2$. In our model, in order to control the value of $\gamma$, edges 
are connected from hubs to non-hub nodes. To reproduce hierarchical networks with $\gamma < 2$, too many 
edges from hubs to non-hub nodes would be required, which increases the complexity of the model. In addition, the fractalness
of the network would be lost in this process.

\bibitem{jeong}  H. Jeong, B. Tombor, R. Albert, Z.N. Oltvai, A.-L. Barab\'{a}si,  Nature {\bf 407}, 651 (2000).

\bibitem{wagner} A. Wagner, D. A. Fell,  Proc. R. Soc. London B {\bf 268}, 1803 (2001).

\bibitem{wuchty2} S. Wuchty, Mol. Biol. Evol. {\bf 18}, 1694 (2001).

\bibitem{wagner2} A. Wagner, Mol. Biol. Evol. {\bf 18}, 1283 (2001).

\bibitem{sitges} A.-L. Barab\'{a}si, Z. Deszo, E. Ravasz, S.H. Yook, and Z. Oltvai {\it(Sitges Proc. on Complex Networks,
2004).}

\bibitem{greek} M. Faloutsos, P. Faloutsos, and C. Faloutsos, Comput. Commun. Rev. {\bf 29}, 251 (1999). 

\bibitem{ourwork} J.C. Nacher, N. Ueda, T. Yamada, M. Kanehisa and T. Akutsu, e-print archive, q-bio.MN/0403045.


\bibitem{comment} Though non-hub nodes with the highest degree in $m$ copies 
are connected to the main hub, we can modify the construction such that
required number of non-hub nodes are connected to the main hub.
In order to maintain the power-law for $P(k)$ and $C(k)$,
it is enough to connect the main hub (at the $i$-th iteration)
with appropriate number of pairs of adjacent non-hub nodes.
In such a case, exponent will be more flexible.
However, fractalness of the network will be lost.














\end{thebibliography}
\end{document}